\documentclass[pra,12pt,tightenlines,showpacs,superscriptaddress]{revtex4}

\usepackage{graphicx}
\usepackage{latexsym}

\begin{document}
\newcommand{\sgn}{\,\mbox{sgn}\,}
\newcommand{\mod}{\,\mbox{mod}\,}
\newcommand{\Fix}{\,\mbox{Fix}\,}
\newcommand{\saw}{\,\mbox{saw}\,}

\title{Hamiltonian mappings and circle packing phase spaces}

\author{A.~J.~Scott}
\email{ajs@maths.uq.edu.au}
\affiliation{Department of Mathematics,}
\author{C.~A.~Holmes}
\affiliation{Department of Mathematics,}
\author{G.~J.~Milburn}
\affiliation{Department of Physics, \\The University of Queensland, QLD
4072 Australia.}

\begin{abstract}
We introduce three area preserving maps with phase space structures which 
resemble circle packings. Each mapping is derived from a kicked Hamiltonian 
system with one of three different phase space geometries (planar, hyperbolic 
or spherical) and exhibits an infinite number of coexisting stable periodic 
orbits which appear to `pack' the phase space with circular resonances.
\end{abstract}

\pacs{05.45.Ac, 05.45.Df \\
MSC numbers: 37J10, 52C26, 30F40 \\ 
Keywords: Nonlinear dynamics; Circle packings}

\maketitle

\section{Introduction}
Nonlinear dynamical systems offer a rich and seemingly endless variety of 
behaviour. In this paper we introduce three Hamiltonian mappings with phase 
space structures which resemble circle packings \cite{bullet,bullet2,herrmann,parker,keen}. 
A circle packing is a set with empty interior and whose complement is the union 
of disjoint open circular discs. Circle packings are often constructed using 
geometrical or group theoretical methods, for example limit sets of Kleinian 
groups \cite{maskit}, but here we may have uncovered them in the context of 
Hamiltonian mechanics. Each of the three mappings acts on a two-dimensional 
manifold with constant curvature, defining one of three different phase space 
geometries: spherical (positive curvature), planar (zero curvature) or 
hyperbolic (negative curvature). In each case an infinite number of stable 
periodic orbits coexist and seem to `pack' the manifold with circular resonances. 
Some analytical data has been found for these maps including the location of a 
large number of periodic orbits, but there remain many open questions, the most 
critical being whether the circular resonances densely pack the phase space. 
That is, whether the residual set not covered by the resonances has empty 
interior. If this is the case then we have indeed found new examples of circle 
packings. In sections II, III, and IV we introduce, respectively, the planar, 
hyperbolic and spherical mappings, and include all analytical findings. Finally, 
in Section V we discuss some open questions concerning these maps.

\section{Planar Map}

The Hamiltonian which generates our mapping on the Euclidean plane is
\begin{displaymath}
H(x,p,t)=\frac{1}{2}\omega(x^2+p^2)+\mu|x|\sum_{n=-\infty}^{\infty}\delta(t-n),
\end{displaymath}
where $x$, $p$ and $t$ are position, momentum and time respectively, and, 
$\omega\in[0,2\pi)$ and $\mu\geq 0$ are parameters. In the planar case $\mu$ may 
be set to unity by rescaling $x$ and $p$. The mapping which takes $(x,p)$ from 
just before a kick to one period later is
\begin{equation}
 \left[\begin{array}{c} x^{n+1} \\ p^{n+1} \end{array}\right] = F \left[\begin{array}{c} x^{n} \\ p^{n} \end{array}\right] = \left[\begin{array}{cc} \cos{\omega} & \sin{\omega} \\ -\sin{\omega} & \cos{\omega}\end{array}\right]\left[\begin{array}{c} x^n\\ p^n-\mu s^n\end{array}\right]
\end{equation}
where $s^n\equiv\sgn x^n$ ($\sgn x$ is the signum function with the convention 
$\sgn 0=0$). Between kicks the Hamiltonian is that of a simple harmonic 
oscillator and all phase space points rotate clockwise about the origin through 
an angle of $\omega$. The effect of the kick is to add a position dependent 
shift in the momentum of $-\mu\sgn x$. If $x\neq 0$ the tangent mapping, 
$\partial F(x,p)/\partial(x,p)$, is simply a linear rotation with eigenvalues 
$e^{\pm i\omega}$. Thus if an orbit is to have nonzero Lyapunov exponents, such 
as an unstable periodic orbit, we need at least one iterate of the orbit on the 
line $x=0$. However, in this instance the tangent mapping is undefined. We 
overcome this problem simply by defining an unstable orbit to be one with a 
point on $x=0$, and all other orbits stable.

If we let $z=x+ip$ then we can rewrite the mapping as
\begin{eqnarray*}
z^{n+1} & = & Fz^n \\
        & = & e^{-i\omega}(z^n-i\mu s^n).
\end{eqnarray*}
Suppose that $z^0$ is a periodic point with period $n$. That is,
\begin{eqnarray*}
 z^0 & = & z^n \\
     & = & e^{-in\omega}z^0-i\mu\sum_{k=0}^{n-1} e^{i(k-n)\omega}s^k.
\end{eqnarray*}
Solving for $z^0$ we obtain
\begin{equation}
z^0=\frac{i\mu}{1-e^{in\omega}}\sum_{k=0}^{n-1} e^{ik\omega}s^k.
\label{perequ}
\end{equation}
If $z^0$ is a stable periodic point then $s^k=\pm 1$, which simply states 
whether the iterate $z^k$ is on the right or left side of the complex plane. The 
unstable periodic orbits must have at least one point on $x=0$. That is, $s^k=0$ 
for some $k=0\dots n-1$. Note that if $n\omega$ is an integer multiple of $2\pi$ 
then equation (\ref{perequ}) is invalid. In these cases one finds that there are 
an infinite number of period $n$ orbits. But when this is not the case, each 
periodic orbit is uniquely determined by the sequences 
$\{s^k=0,\pm 1\}_{k=0\dots n-1}$, with at most two $s^k$ being zero (see the 
Lemma below). By cycling a particular sequence we obtain the $n$ iterates of the 
periodic orbit. Although each periodic orbit is uniquely represented by a 
sequence, not every sequence represents a periodic orbit. Hence, we still need 
to find which sequences are legitimate. This task can be simplified by noting 
that $F$ can be decomposed into the two involutions $I_1$ and $I_2$, where
\begin{eqnarray*}
I_1z & = & -z^*-i\mu\sgn x \\
I_2z & = & -e^{-i\omega}z^* \\
{I_1}^2 & = & {I_2}^2 = 1 \\
I_1FI_1 & = &I_2FI_2 =F^{-1}
\end{eqnarray*}
and 
\begin{displaymath}
F=I_2\circ I_1.
\end{displaymath}
It is well-known that whenever a mapping admits such a decomposition there exist 
symmetric periodic orbits having points which lie on the fixed lines of each 
involution \cite{pina}. The fixed line of an involution $I$ is the set
\begin{displaymath}
\Fix(I)\equiv\{z|Iz=z\}.
\end{displaymath} 
{\bf Theorem} \cite{lamb}: Let $F$ be an invertible mapping and $I$ an 
involution ($I^2=1$) satisfying $IFI=F^{-1}$. Let $z\in\Fix(I)$ and suppose that 
$z$ is a periodic point of $F$ with least period $n$. Then $\Fix(I)$ contains 
no other points of the periodic orbit if $n$ is odd, or exactly one other, 
$F^{n/2}z$, if $n$ is even.    
\vspace{.1in}\\
Consider the unstable periodic orbits. From our definition, an unstable orbit 
must have at least one point lying on $x=0$. Noting that $\Fix(I_1)=\{x+ip|x=0\}$ 
and putting $I=I_1$ in the above theorem we obtain the following Lemma:
\vspace{.1in}\\
{\bf Lemma :} Let $\{s^k\}_{k=0\dots n-1}$ be the sequence defining an unstable 
periodic orbit of least period $n$. If $n$ is odd then the sequence contains 
exactly one zero, $s^0=0$ say, and if $n$ is even the sequence contains exactly 
two zeros, $s^0=0$ and $s^{n/2}=0$.
\vspace{.1in}

We now know that all unstable periodic orbits have one or two points on the 
fixed line of the involution $I_1$ ($x=0$). Many of the stable periodic orbits 
have points on the fixed line of the second involution $I_2$, which is given by 
the equation $x=p\tan{\omega/2}$. In this case one can also show that the orbits 
with odd period have one point on this line while the orbits with even period 
have two. One cannot show, however, that the stable periodic orbits must have 
point on this line, indeed, some do not. There is a second decomposition of the 
mapping, $F=\tilde{I}_2\circ \tilde{I}_1$, where $\tilde{I}_1=-I_1$ has the 
fixed line $p=\mu/2\sgn x$ and $\tilde{I}_2=-I_2$ has the fixed line 
$p=-x\tan{\omega/2}$. We are unsure as to whether every stable periodic orbit 
has a point on one of the above fixed lines. However, numerical 
investigations seem to suggest that they do. For the case $\omega=\pi(\sqrt{5}-1)$ 
we have checked all $2^{41}-2$ of
the possible sequences for a stable periodic orbit of period $\leq 40$ and 
found only those with points on these lines. Hence we tentatively conjecture
that all stable periodic orbits have one or two points on at least one of the 
fixed lines of $I_1$, $\tilde{I}_1$ and $\tilde{I}_2$.
\vspace{.1in}\\
{\bf Proposition :} If $\{s^k\}_{k=0\dots n-1}$ is the sequence defining a 
periodic orbit of least period $n$ with $z^0$ being the first point, then
\begin{enumerate}
\item $s^k=-s^{n-k}$ if $z^0\in\Fix(I_1)$,
\item $s^k=-s^{n-k-1}$ if $z^0\in\Fix(I_2)$,
\item $s^k=s^{n-k}$ if $z^0\in\Fix(\tilde{I}_1)$,
\item $s^k=s^{n-k-1}$ if $z^0\in\Fix(\tilde{I}_2)$,
\end{enumerate}
for $k=1\dots n-1$.
\vspace{.1in}
\\{\bf Proof :} We will only prove the first case. The others are similar. We 
know that $F^nz^0=z^0$ and $I_1z^0=z^0$, since $z^0\in\Fix(I_1)$. The iterate 
$z^k$, $1\leq k<n$, is another point of the periodic orbit and
\begin{eqnarray*}
I_1z^k & = & I_1F^{k}z^0 \\
       & = & I_1I_1F^{-k}I_1z^0 \\
       & = & F^{-k}z^0 \\
       & = & F^{n-k}z^0 \\
       & = & z^{n-k}.
\end{eqnarray*}
Thus $I_1z^k$ is also a point of the periodic orbit, and since $I_1$ changes 
the sign of $x$ the above equation gives the formula $s^k=-s^{n-k}$. \hspace{.1in}$\Box$
\vspace{.1in}

\begin{figure}[p]
\includegraphics[scale=1]{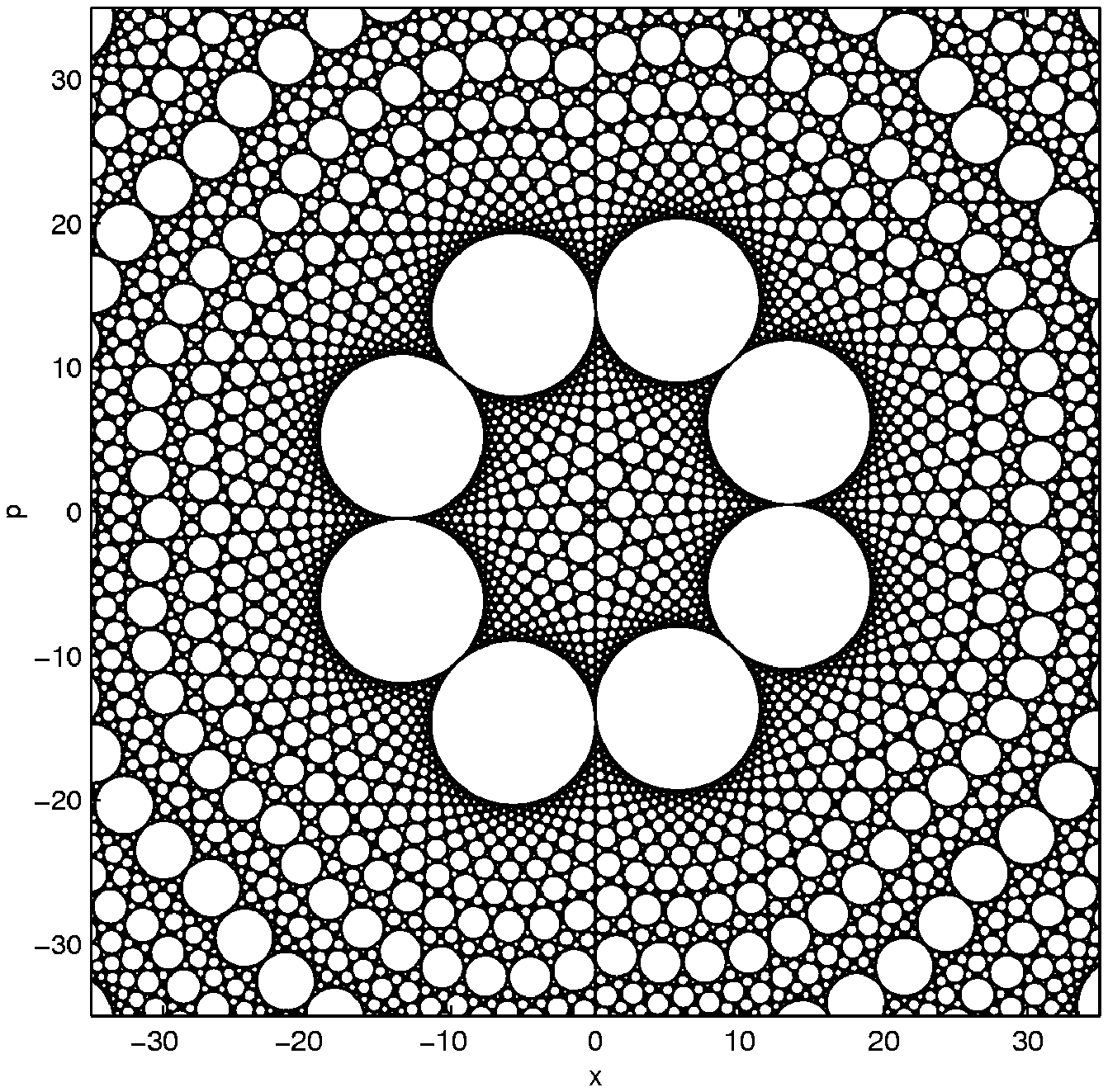}
\caption{The phase space portrait for the planar map when $\omega=\pi(\sqrt{5}-1)$ and $\mu=1$.}
\label{fig1}
\end{figure}

The phase space portrait for $\omega=\pi(\sqrt{5}-1)$ and $\mu=1$ is shown in 
Fig. \ref{fig1}. The unstable orbits are in black and consist of unstable periodic and 
aperiodic motion. They were generated by plotting all images and preimages of 
the vertical line $x=0$. The circular discs are resonances and are packed 
entirely with stable orbits. At the center of each of these lies a stable 
periodic orbit. The positions of a large number of periodic orbits have been 
found analytically. For want of a better term, we will call these `first order' 
periodic orbits. All of the 
resonances visible in Fig. \ref{fig1} are of first order. Only on greater levels of 
magnification does one encounter resonances of higher order.  
   
The planar mapping has a stable first order periodic orbit of period $n$ if 
there exists an integer $m$ such that $1\leq m\leq n$, $\gcd (n,m)=1$, and
\begin{displaymath}
\frac{m-1}{n}<\frac{\omega}{2\pi}<\frac{m}{n}\quad\mbox{if $n$ is even,}
\end{displaymath}
or
\begin{displaymath}
\frac{m-1/2}{n}<\frac{\omega}{2\pi}<\frac{m}{n}\quad\mbox{if $n$ is odd.}
\end{displaymath} 
These orbits are born near the origin for the smaller value of $\omega$ and then 
are destroyed at infinity when $\omega/2\pi=m/n$. The sequences defining these 
orbits are 
\begin{equation}
s^k=\sgn\left(\sin{2\pi\frac{km+1/4}{n}}\right)
\label{stables}
\end{equation}
$k=0\dots n-1$. The $n$ iterates of each periodic orbit are found by inserting 
the $n$ cycles of the corresponding sequence into (\ref{perequ}). The orbits 
with odd period come in pairs. The second of each pair is found by negating 
every term of the sequence. All these orbits, apart from those of period 1 and 2, 
are created with a corresponding unstable periodic orbit. These unstable orbits 
exist for the same $\omega$ as the stable and have sequences
\begin{equation}
s^k=\sgn\left(\sin{2\pi\frac{km}{n}}\right).
\label{unstables}
\end{equation} 
There is also an unstable period 1 orbit at the origin which exists for all 
$\omega$. 

The first order periodic orbits were found by observing that at infinity the 
kick has no effect and the map is simply a linear rotation through an angle of 
$\omega$. Consequently, when $\omega/2\pi=m/n$ there will be period $n$ orbits 
at infinity and these orbits must have sequences given by (\ref{stables}) or 
(\ref{unstables}), depending on their stability. Further investigation reveals 
that they were born near the origin at a smaller value of $\omega$ when points 
of the stable periodic orbit intersected the line $x=0$. Hence one can use 
(\ref{perequ}) to find out when this occurred and obtain the above results. 

\begin{figure}[p]
\includegraphics[scale=1]{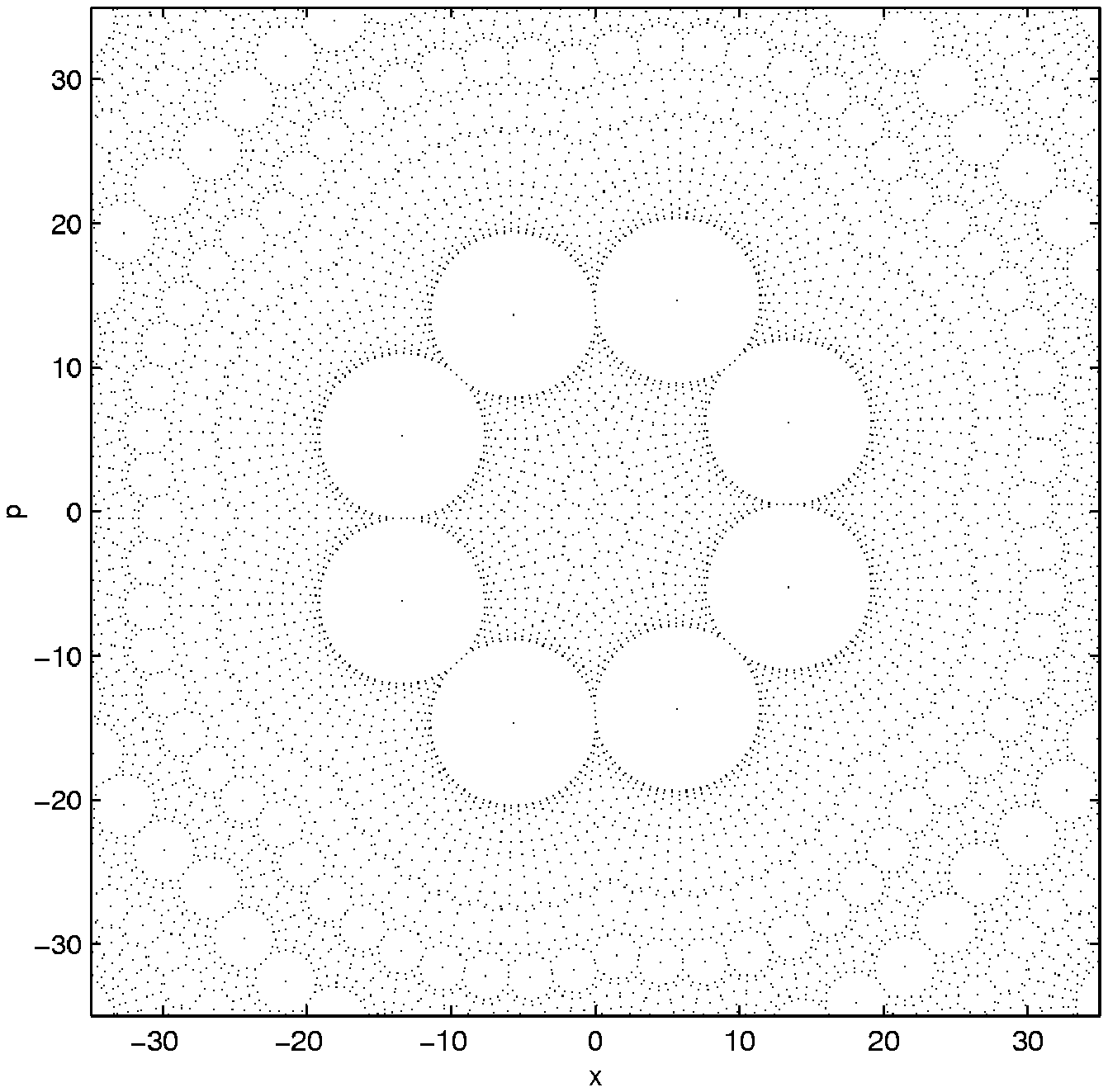}
\caption{The stable first order periodic orbits in Fig. \ref{fig1}.}
\label{fig2}
\end{figure}

\begin{figure}[p]
\includegraphics[scale=1]{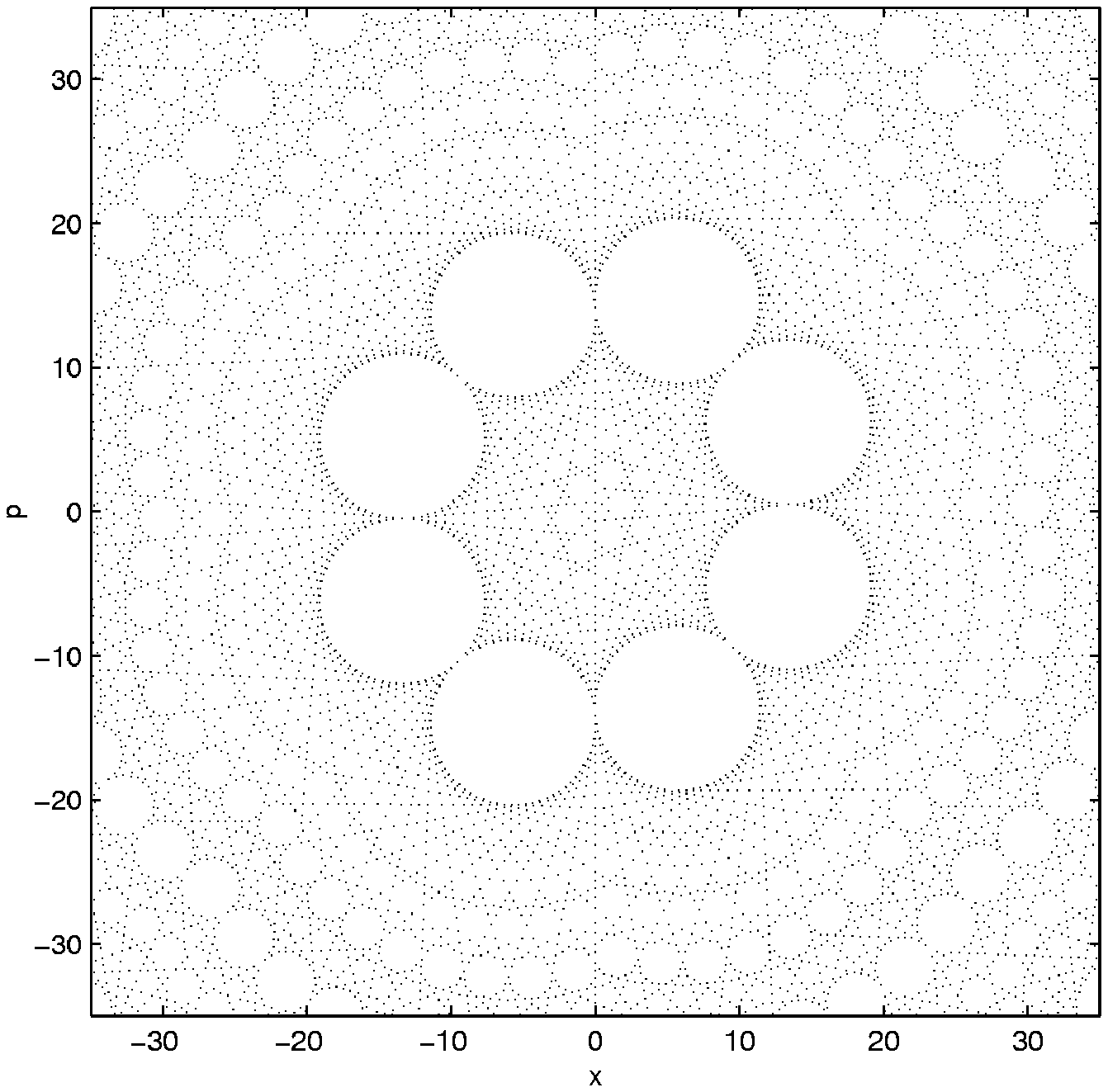}
\caption{The unstable first order periodic orbits in Fig. \ref{fig1}.}
\label{fig3}
\end{figure}

All these orbits revolve around the origin under iteration, and in general, the 
orbits for which $m/n$ is larger are closer to the origin since they are 
destroyed at infinity for a greater value of $\omega$. In Fig. \ref{fig2} and Fig. \ref{fig3} we 
have plotted, respectively, the positions of all stable and unstable first order 
periodic orbits. The values of $\omega$ and $\mu$ are the same as in Fig. \ref{fig1}. The 
radii of the circular resonances enclosing each of the stable first order 
periodic orbits can also been found. This may be done by noting that every 
stable periodic orbit must have an iterate with its circular resonance tangent 
to the line $x=0$. The radii is then given by the $x$ position of that iterate. 
These reduce to
\begin{displaymath}
r_n=\frac{1}{2}\tan{\frac{\omega n}{4}}\quad\mbox{if $n$ is even,}
\end{displaymath}
or
\begin{displaymath}
r_n=\frac{1}{2}\tan{\left(\frac{\omega n}{2}+\frac{\pi}{2}\right)}\quad\mbox{if $n$ is odd.}
\end{displaymath}
All iterates of an orbit inside a resonance can be written as
\begin{equation}
z^k=z_p^{k\,\text{mod}\, n}+(z^0-z_p^0)e^{-ik\omega} \label{localz}
\end{equation}
where $z_p^0$ is the period $n$ point at the center of the resonance containing 
$z^0$, and
\begin{displaymath}
|z^0-z_p^0|<r_n.
\end{displaymath}
If $\omega/2\pi$ is irrational then the orbit iterates in a quasiperiodic 
circular motion about the central periodic orbit. However, when $\omega/2\pi$ 
is rational, that is, $\omega/2\pi=p/q$ where $p$ and $q$ are integers with 
$\gcd (p,q)=1$, then $z^{nq}=z^0$. Hence the orbit is periodic. In this case 
each image of the vertical line $x=0$ is a collection of parallel line segments 
at one of only $q$ different possible angles. Consequently, the resonances are 
all polygons which tile the phase plane. The phase space portrait for 
$\omega/2\pi=5/8$ and $\mu=1$ is shown in Fig. \ref{fig4}. Here the phase plane is tiled 
with octagons of decreasing size. In Fig. \ref{fig5} where $\omega/2\pi=3/5$ the phase 
plane is tiled with pentagons and decagons.

\section{Hyperbolic Map}

The hyperbolic map is derived from the Hamiltonian
\begin{displaymath}
H({\bf K},t)=\omega K_3+\mu|K_1|\sum_{n=-\infty}^{\infty}\delta(t-n),
\end{displaymath}
where ${\bf K} =(K_1,K_2,K_3)=(-x_2p_3-x_3p_2,x_3p_1+x_1p_3,x_1p_2-x_2p_1)$ is 
the Minkowski 3-vector for a particle confined to a pseudosphere\cite{balazs}, 
normalized such that ${\bf K}$ lies on the hyperboloid
\begin{equation}
1+{K_1}^2+{K_2}^2={K_3}^2,\quad K_3>0.
\label{knorm}
\end{equation}
The evolution of ${\bf K}$ under the above Hamiltonian is governed by the 
equations
\begin{displaymath}
\dot{K_i}=\{K_i,H\},\quad \{K_1,K_2\}=-K_3,\quad \{K_2,K_3\}=K_1,\quad \{K_3,K_1\}=K_2,
\end{displaymath}
where $\{\cdot\,,\cdot\}$ are the Poisson brackets. The mapping which takes 
${\bf K}$ from just before a kick to one period later is
\begin{eqnarray*}
\left[\begin{array}{c} K_1^{n+1} \\ K_2^{n+1} \\ K_3^{n+1} \end{array}\right] = F\left[\begin{array}{c} K_1^n \\ K_2^n \\ K_3^n \end{array}\right] & = & \left[\begin{array}{ccc} \cos{\omega} & -\sin{\omega} & 0 \\ \sin{\omega} & \cos{\omega} & 0 \\ 0 & 0 & 1 \end{array}\right]\left[\begin{array}{ccc} 1 & 0 & 0 \\ 0 & \cosh{\mu s^n} & \sinh{\mu s^n} \\ 0 & \sinh{\mu s^n} & \cosh{\mu s^n} \end{array}\right]\left[\begin{array}{c} K_1^n \\ K_2^n \\ K_3^n\end{array}\right] \\
& \equiv & \mbox{F}(s^n){\bf K}^n
\end{eqnarray*}
where $s^n\equiv\sgn K_1^n$. 

The unstable orbits are defined to be those with at least one point with $K_1=0$. 
In a similar fashion to the planar case we can label all periodic orbits with 
sequences $\{s^k=0,\pm 1\}_{k=0\dots n-1}$. Then the position of the periodic 
point must be the solution of
\begin{equation}
{\bf K}=\mbox{F}(s^{n-1})\mbox{F}(s^{n-2})\dots \mbox{F}(s^0){\bf K},
\label{perhequ}
\end{equation}
which lies on the hyperboloid (\ref{knorm}). By using the previous theorem and 
decomposing $F$ into the two involutions,
\begin{eqnarray*}
I_1{\bf K} & = & (-K_1,K_2\cosh(\mu\sgn K_1)+K_3\sinh(\mu\sgn K_1),K_2\sinh(\mu\sgn K_1)+K_3\cosh(\mu\sgn K_1)), \\
I_2{\bf K} & = & (-K_1\cos{\omega}-K_2\sin{\omega},-K_1\sin{\omega}+K_2\cos{\omega},K_3),
\end{eqnarray*}
where $F=I_2\circ I_1$, we can show that the lemma and proposition in the 
previous section are also true in the hyperbolic case. The second pair of 
involutions for the hyperbolic map are
\begin{eqnarray*}
\tilde{I}_1{\bf K} & = & (K_1,-K_2\cosh(\mu\sgn K_1)-K_3\sinh(\mu\sgn K_1),K_2\sinh(\mu\sgn K_1)+K_3\cosh(\mu\sgn K_1)), \\
\tilde{I}_2{\bf K} & = & (K_1\cos{\omega}+K_2\sin{\omega},K_1\sin{\omega}-K_2\cos{\omega},K_3),
\end{eqnarray*}
with $F=\tilde{I}_2\circ \tilde{I}_1$. Hence the mapping has symmetry lines on 
the hyperboloid (\ref{knorm}) of $K_1=0$, $K_1=-K_2\tan{\omega/2}$, 
$K_2=-K_3\tanh(\mu/2\sgn K_1)$, and $K_2=K_1\tan{\omega/2}$ which are the fixed 
lines of the involutions $I_1$, $I_2$, $\tilde{I}_1$, and $\tilde{I}_2$, 
respectively.

To display the hyperbolic phase space we will use the conformal disk model 
\cite{ratcliffe} where all angles are Euclidean. Hence all circles will look 
like Euclidean circles, but some will appear larger than others of the same 
periodic orbit because the model distorts distances. The mapping  
\begin{displaymath}
x=\frac{K_1}{1+K_3},\quad y=\frac{K_2}{1+K_3},
\end{displaymath}
takes all points on the hyperboloid (\ref{knorm}) into the disk $x^2+y^2<1$ in 
the Euclidean plane. The phase space portraits for $\omega=\pi(\sqrt{5}-1)$, 
$\mu=0.04$ and $\mu=0.2$ are shown in Fig. \ref{fig6} and Fig. \ref{fig7}, respectively. On 
comparing Fig. \ref{fig6} with Fig. \ref{fig1} we can see that the hyperbolic map resembles the 
planar map when $\mu$ is small. In fact, under a suitable transformation of 
variables the hyperbolic map can be approximated by the planar map when close 
to the origin and $\mu$ is small enough. Consequently, the first order periodic 
orbits of the hyperbolic map exist for the same $\omega$ as in the planar map 
with the exception that they are destroyed at infinity ($x^2+y^2$=1) at a 
smaller value of $\omega$ dependent on $\mu$. The hyperbolic mapping has a 
stable first order periodic orbit of period $n$ if there exists an integer $m$ 
such that $1\leq m\leq n$, $\gcd (n,m)=1$, and
\begin{displaymath}
\frac{m-1}{n}<\frac{\omega}{2\pi}<f_{mn}(\mu)\leq\frac{m}{n}\quad\mbox{if $n$ is even,}
\end{displaymath}
or
\begin{displaymath}
\frac{m-1/2}{n}<\frac{\omega}{2\pi}<f_{mn}(\mu)\leq\frac{m}{n}\quad\mbox{if $n$ is odd.}
\end{displaymath} 
The sequences defining these orbits are given by (\ref{stables}), and the 
$k$-th iterate is found by inserting the $k$-th cycle of the sequence into 
(\ref{perhequ}) and solving on the hyperboloid (\ref{knorm}). Again, the orbits 
with odd period come in pairs. The second of each pair is found by negating the 
sequence. The corresponding unstable orbits have sequences given by 
(\ref{unstables}). The functions $f_{mn}(\mu)$ could not be found analytically 
except in the special cases of $(m,n)=(1,1)$ where
\begin{displaymath}
f_{11}(\mu)=1-\frac{1}{\pi}\arcsin\left(\tanh{\frac{\mu}{2}}\right),
\end{displaymath}
and $(m,n)=(1,2),(1,4),(3,4)$ where
\begin{displaymath}
f_{mn}(\mu)=\frac{m}{n}-\frac{1}{\pi}\arcsin\left(\sin{\frac{\pi}{n}}\tanh{\frac{\mu}{2}}\right).
\end{displaymath}
These can be found by explicitly solving (\ref{perhequ}) for the position of 
each periodic orbit. A much simpler way to find these orbits of low period is 
to make use of the fact that some iterates will lie on the symmetry lines. 
Although the functions $f_{mn}(\mu)$ are generally unknown, the task of finding 
all first order periodic orbits is still quite simple.  One simply replaces the 
inequality $\omega/2\pi<f_{mn}(\mu)$ with the weaker inequality 
$\omega/2\pi<m/n$ together with the condition that the nontrivial solutions of 
(\ref{perhequ}) (with (\ref{stables}) or (\ref{unstables})) have 
${K_3}^2-{K_1}^2-{K_2}^2=t^2>0$ (the case of $f_{mn}(\mu)<\omega/2\pi<m/n$ 
corresponds to ${K_3}^2-{K_1}^2-{K_2}^2=-t^2$).

In the previous section we found that when $\omega/2\pi$ is rational the phase 
plane of the planar map is entirely filled with polygons. This does not 
occur in the hyperbolic case. The tangent mapping to the planar map is the same 
linear rotation matrix for both the left and right half of the phase plane. 
Consequently, the local rotation about every stable periodic orbit is $-\omega$ 
(see Eq. (\ref{localz})), and if $\omega/2\pi$ is rational then this rotation is 
itself periodic. The hyperbolic map, however, has different tangent mappings for 
the left and right half of the hyperbolic plane. Hence the local rotation about  
each stable periodic orbit will be different. However we can still choose 
$\omega$ and $\mu$ such that the rotation about one particular orbit is periodic. 
When this occurs we find that each resonance of the periodic orbit forms a polygon. 
For example, by choosing $\mu=0.5$ and setting 
$\omega=\pi+\arccos\tanh^2\mu/2=4.652..$ we find that the resonances enclosing 
each of the two period 1 orbits are hyperbolic squares (see Fig. \ref{fig8}). 
Each point inside these squares is a period 4 orbit, rotating clockwise exactly
$\pi/2$ radians at each iteration. 

\section{Spherical Map}

The spherical map is given by the Hamiltonian
\begin{displaymath}
H({\bf J},t)=\omega J_3+\mu|J_1|\sum_{n=-\infty}^{\infty}\delta(t-n),
\end{displaymath}
where $({\bf J})_i=J_i=\epsilon_{ijk}x_jp_k$ $(i=1,2,3)$ are the three 
components of angular momentum for a particle confined to a sphere, normalized 
such that
\begin{equation}
{J_1}^2+{J_2}^2+{J_3}^2=1.
\label{jnorm}
\end{equation}
The evolution of ${\bf J}$ under the above Hamiltonian is governed by the 
equations
\begin{displaymath}
\dot{J_i}=\{J_i,H\},\quad \{J_i,J_j\}=\epsilon_{ijk}J_k.
\end{displaymath}
The mapping which takes ${\bf J}$ from just before a kick to one period later 
is
\begin{eqnarray}
{\bf J}^{n+1} = \left[\begin{array}{c} J_1^{n+1} \\ J_2^{n+1} \\ J_3^{n+1} \end{array}\right] & = & \left[\begin{array}{ccc} \cos{\omega} & -\sin{\omega} & 0 \\ \sin{\omega} & \cos{\omega} & 0 \\ 0 & 0 & 1 \end{array}\right]\left[\begin{array}{ccc} 1 & 0 & 0 \\ 0 & \cos{\mu s^n} & -\sin{\mu s^n} \\ 0 & \sin{\mu s^n} & \cos{\mu s^n} \end{array}\right]\left[\begin{array}{c} J_1^n \\ J_2^n \\ J_3^n\end{array}\right] \label{jmap}\\
& \equiv & \mbox{F}(s^n){\bf J}^n \nonumber
\end{eqnarray}
where $s^n\equiv\sgn J_1^n$.

The unstable orbits are defined to be those with at least one point for which 
$J_1=0$. Again, we can label all periodic orbits with sequences 
$\{s^k=0,\pm 1\}_{k=0\dots n-1}$. Then the position of the periodic point must 
be the solution of
\begin{equation}
{\bf J}=\mbox{F}(s^{n-1})\mbox{F}(s^{n-2})\dots \mbox{F}(s^0){\bf J},
\label{persequ}
\end{equation}
on the unit sphere (\ref{jnorm}) with $\sgn J_1=s^0$. By decomposing $F$ into 
the two involutions,
\begin{eqnarray*}
I_1{\bf J} & = & (-J_1,J_2\cos(\mu\sgn J_1)-J_3\sin(\mu\sgn J_1),J_2\sin(\mu\sgn J_1)+J_3\cos(\mu\sgn J_1)), \\
I_2{\bf J} & = & (-J_1\cos{\omega}-J_2\sin{\omega},-J_1\sin{\omega}+J_2\cos{\omega},J_3),
\end{eqnarray*}
where $F=I_2\circ I_1$, we can show that the lemma and proposition of Section II 
hold in this case. The second pair of involutions for the spherical map are
\begin{eqnarray*}
\tilde{I}_1{\bf J} & = & (J_1,-J_2\cos(\mu\sgn J_1)+J_3\sin(\mu\sgn J_1),J_2\sin(\mu\sgn J_1)+J_3\cos(\mu\sgn J_1)), \\
\tilde{I}_2{\bf J} & = & (J_1\cos{\omega}+J_2\sin{\omega},J_1\sin{\omega}-J_2\cos{\omega},J_3),
\end{eqnarray*}
with $F=\tilde{I}_2\circ \tilde{I}_1$. Hence the mapping has symmetry lines on 
the unit sphere (\ref{jnorm}) of $J_1=0$, $J_1=-J_2\tan{\omega/2}$, 
$J_2=J_3\tan(\mu/2\sgn J_1)$, and $J_2=J_1\tan{\omega/2}$ which are the fixed 
lines of the involutions $I_1$, $I_2$, $\tilde{I}_1$, and $\tilde{I}_2$, 
respectively.

We will only display the eastern hemisphere $(J_1>0)$ of the spherical phase 
space. Under the transformation $(J_1,J_2,J_3)\rightarrow (-J_1,-J_2,J_3)$ the 
mapping (\ref{jmap}) remains invariant, hence the phase space structure in the 
the western hemisphere will be symmetrical to the eastern under this reflection. 
The mapping  
\begin{displaymath}
x=\frac{J_2}{1+J_1},\quad y=\frac{J_3}{1+J_1},
\end{displaymath}
takes all points on the eastern hemisphere into the disk $x^2+y^2<1$ in the 
Euclidean plane. All circles in the spherical phase space are transformed to 
Euclidean circles, but some will appear larger than others of the same periodic 
orbit because the mapping distorts distances. The phase space portraits for 
$\omega=\pi(\sqrt{5}-1)$, $\mu=0.02$ and $\mu=\pi(\sqrt{5}-1)$ are shown in 
Fig. \ref{fig9} and Fig. \ref{fig10}, respectively. For small $\mu$ the southern hemisphere can be 
approximated by the planar map, while the northern hemisphere can be 
approximated by the planar map under the transformation 
$\omega\rightarrow 2\pi-\omega$ (compare Fig. \ref{fig9} with Fig. \ref{fig1}). Hence we would 
expect there to be a stable first order periodic orbit of period $n$ if there 
exists an integer $m$ such that $1\leq m\leq n$, $\gcd (n,m)=1$, and
\begin{displaymath}
\frac{m-1}{n}<\frac{\omega}{2\pi}<\frac{m}{n}\quad\mbox{or}\quad\frac{m-1}{n}<1-\frac{\omega}{2\pi}<\frac{m}{n}\quad\mbox{if $n$ is even,}
\end{displaymath}
or
\begin{displaymath}
\frac{m-1/2}{n}<\frac{\omega}{2\pi}<\frac{m}{n}\quad\mbox{or}\quad\frac{m-1/2}{n}<1-\frac{\omega}{2\pi}<\frac{m}{n}\quad\mbox{if $n$ is odd,}
\end{displaymath}
together with some condition on $\mu\in[0,2\pi)$. Again, the sequences defining 
these orbits are given by (\ref{stables}), while the corresponding unstable 
orbits have sequences given by (\ref{unstables}). The position of each periodic 
orbit is found by solving (\ref{persequ}). The condition on $\mu$ for the 
existence of each periodic orbit could not be found analytically except in the 
special cases of $n=1,2,3,4,6.$ The orbits of period $n=1,2,4$ exist for all 
$\mu$ (and $\omega$). The period $3$ orbits exist for
\begin{displaymath}
\cos{\mu}>\frac{\cos{\omega}}{1-\cos{\omega}}
\end{displaymath} 
while the period $6$ orbits exist for
\begin{displaymath}
\cos{\mu}>\frac{-\cos{\omega}}{1+\cos{\omega}}.
\end{displaymath}   
For orbits of higher period one can simply just ignore any condition imposed on 
$\mu$. After solving (\ref{persequ}) we then need to check whether we have 
actually found a periodic orbit. All first order periodic orbits can be found 
in this manner. As for the hyperbolic mapping, some choices of $\omega$ and 
$\mu$ will produce polygonal resonances. 

\section{Discussion and Conclusion}

\begin{figure}[p]
\includegraphics[scale=1]{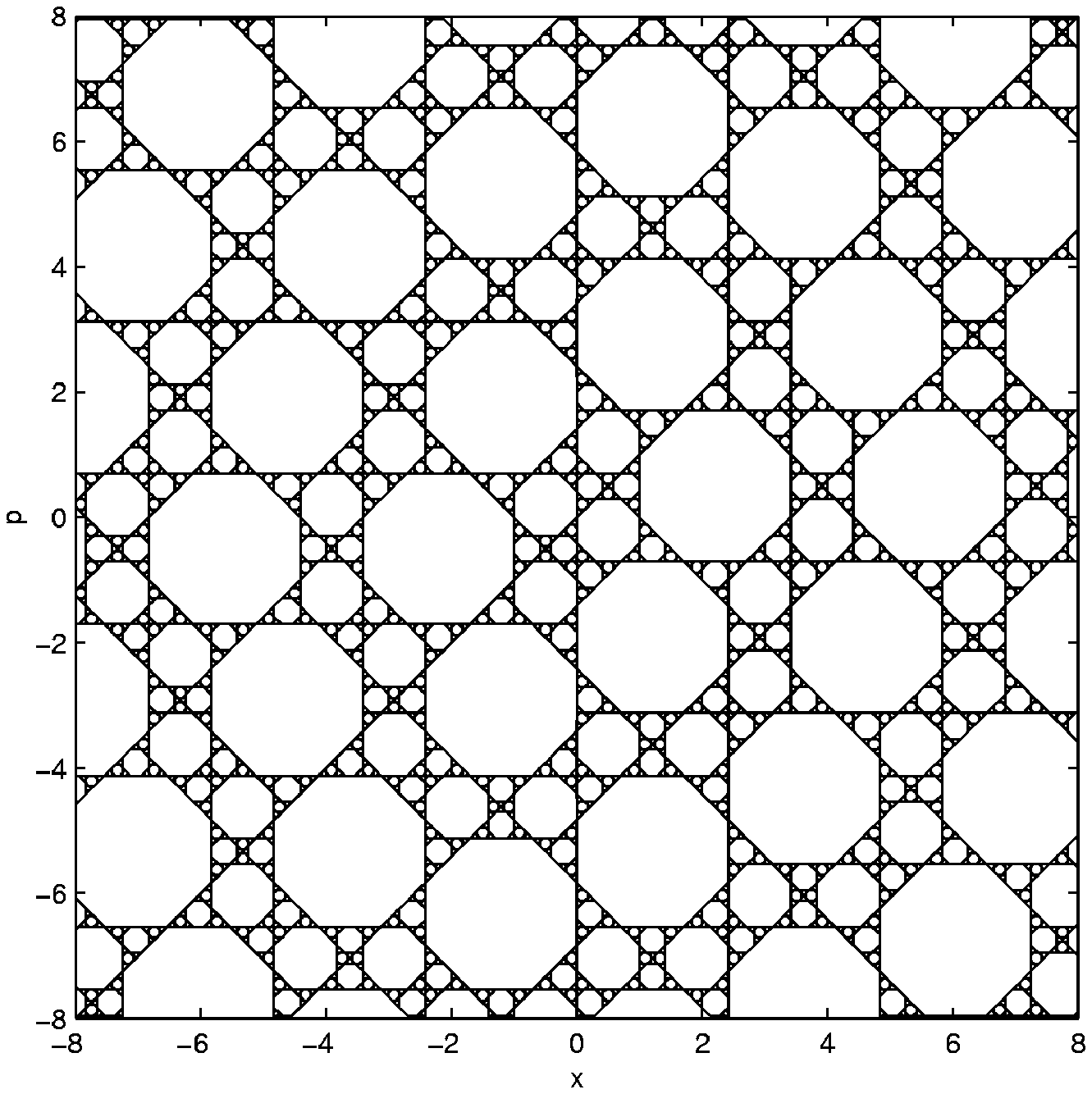}
\caption{The phase space portrait for the planar map when $\omega/2\pi=5/8$ and $\mu=1$.}
\label{fig4}
\end{figure}

\begin{figure}[p]
\includegraphics[scale=1]{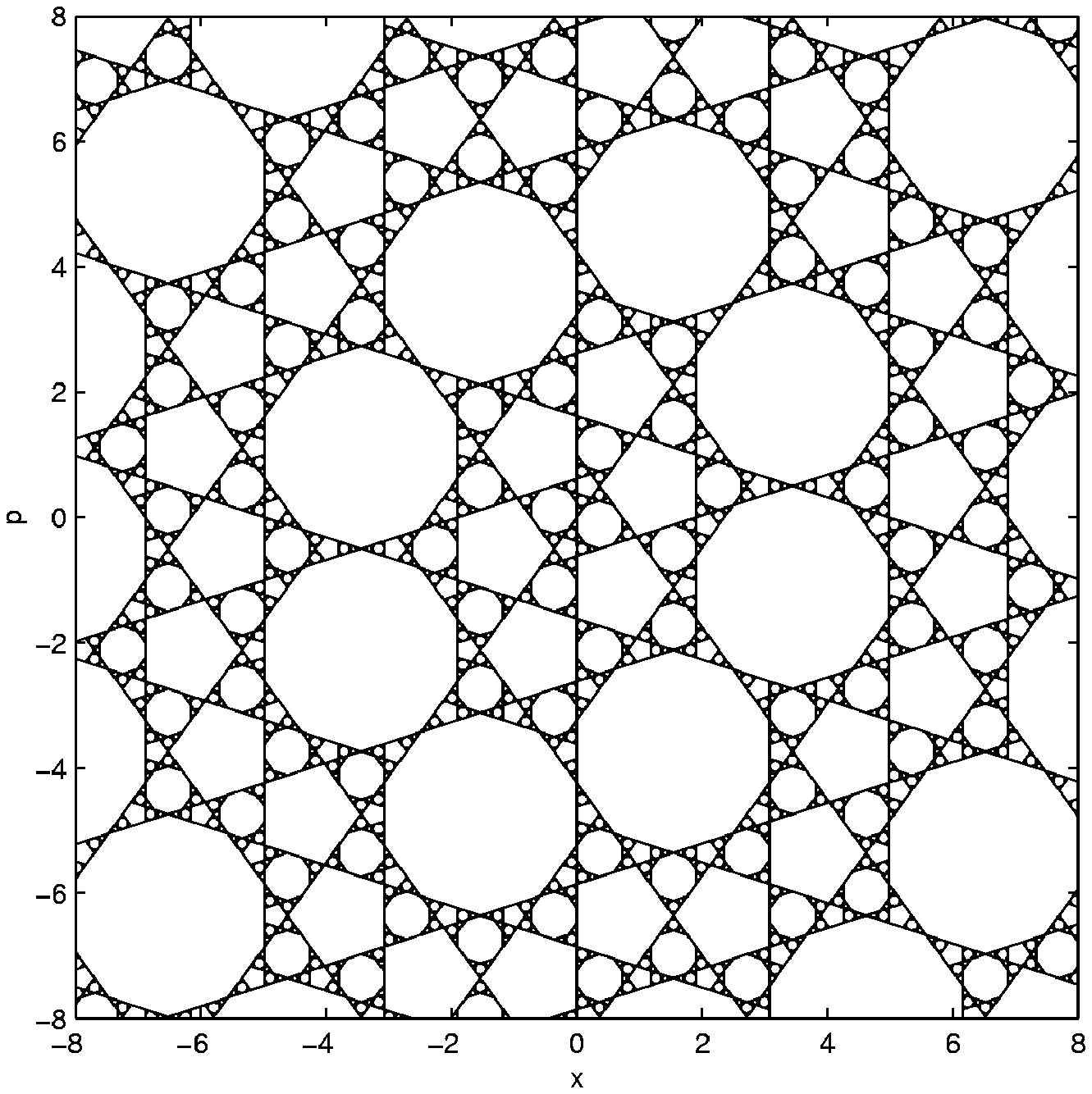}
\caption{The phase space portrait for the planar map when $\omega/2\pi=3/5$ and $\mu=1$.}
\label{fig5}
\end{figure}

\begin{figure}[p]
\includegraphics[scale=1]{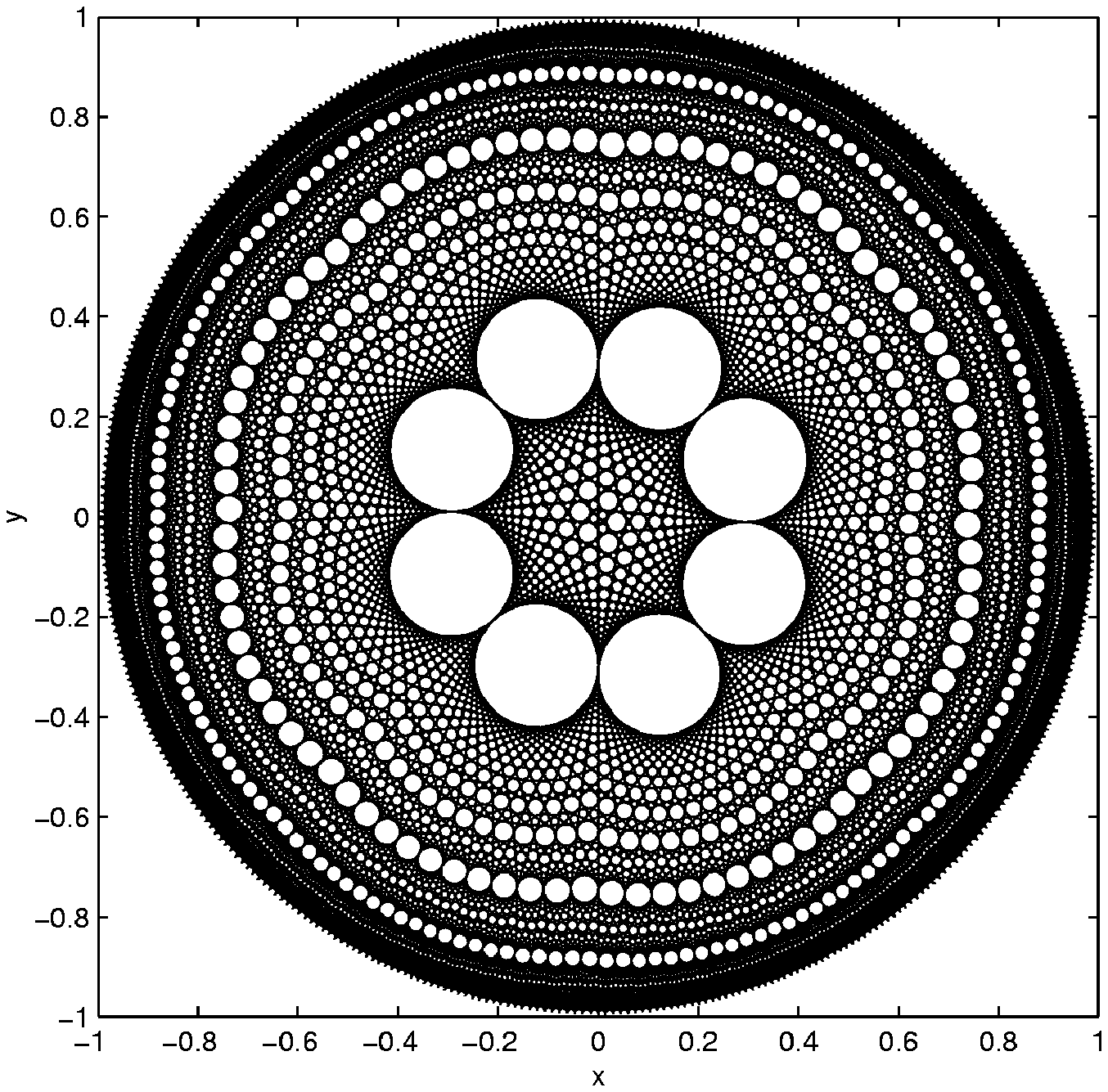}
\caption{The phase space portrait for the hyperbolic map when $\omega=\pi(\sqrt{5}-1)$ and $\mu=0.04$.}
\label{fig6}
\end{figure}

\begin{figure}[p]
\includegraphics[scale=1]{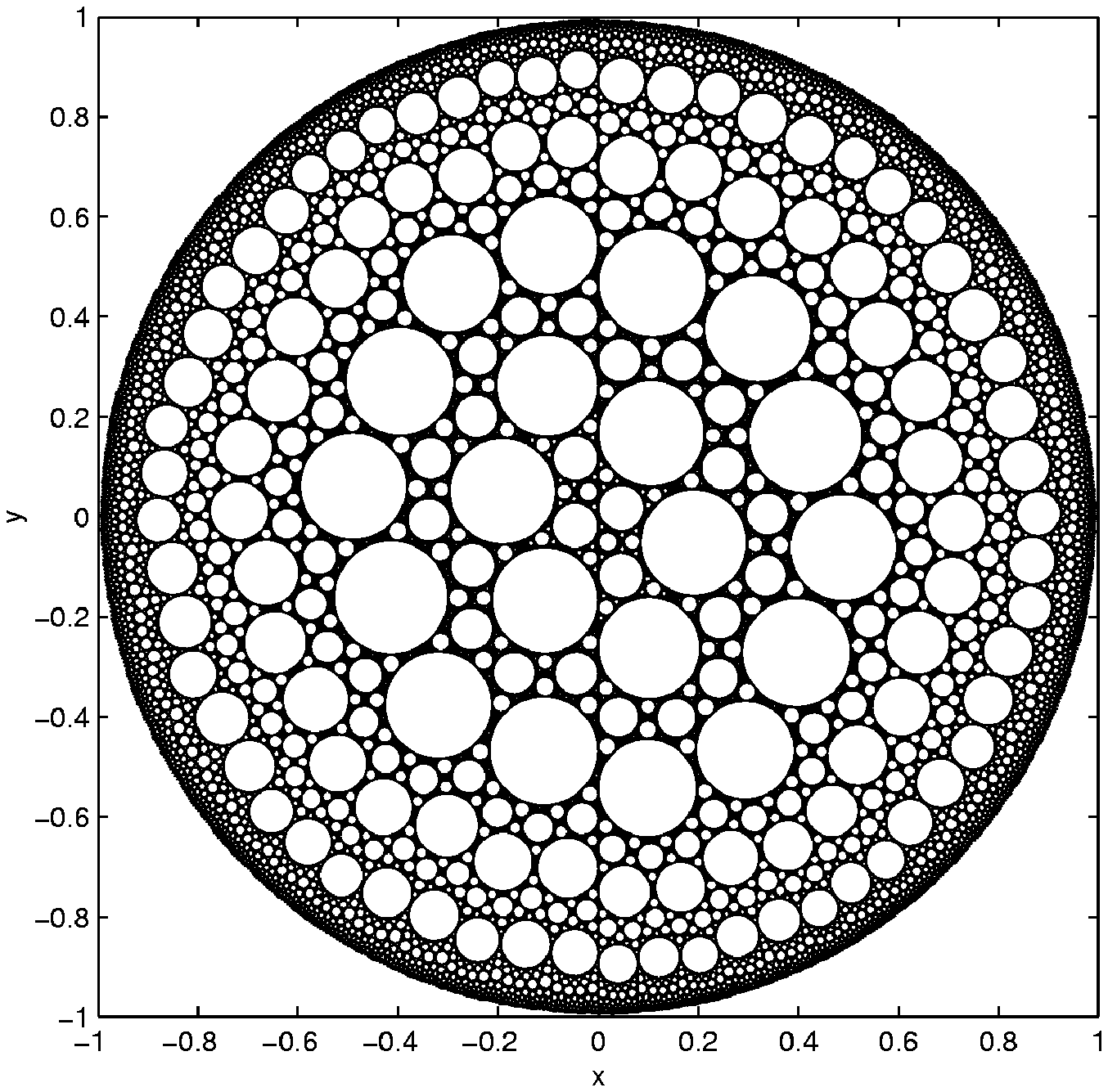}
\caption{The phase space portrait for the hyperbolic map when $\omega=\pi(\sqrt{5}-1)$ and $\mu=0.2$.}
\label{fig7}
\end{figure}

\begin{figure}[p]
\includegraphics[scale=1]{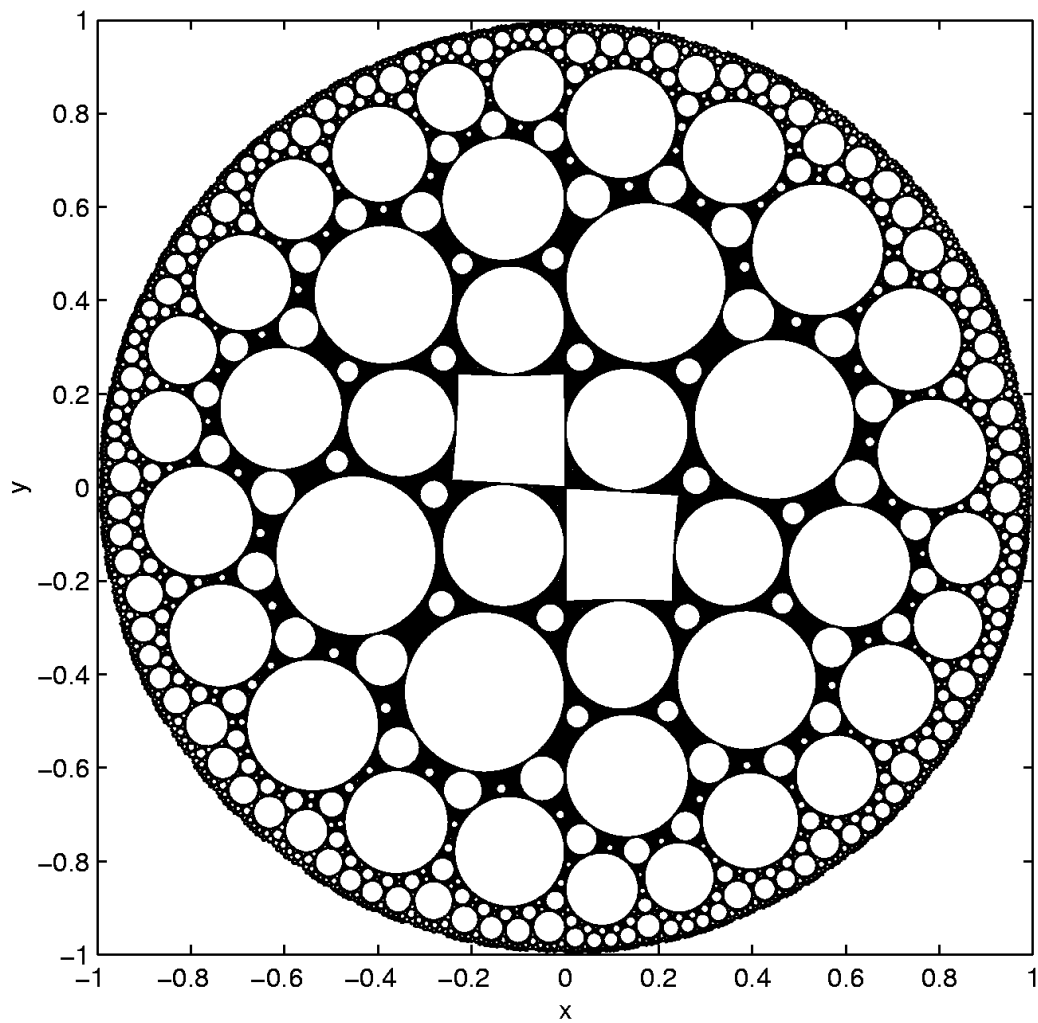}
\caption{The phase space portrait for the hyperbolic map when $\omega=4.652..$ and $\mu=0.5$.}
\label{fig8}
\end{figure}

\begin{figure}[p]
\includegraphics[scale=1]{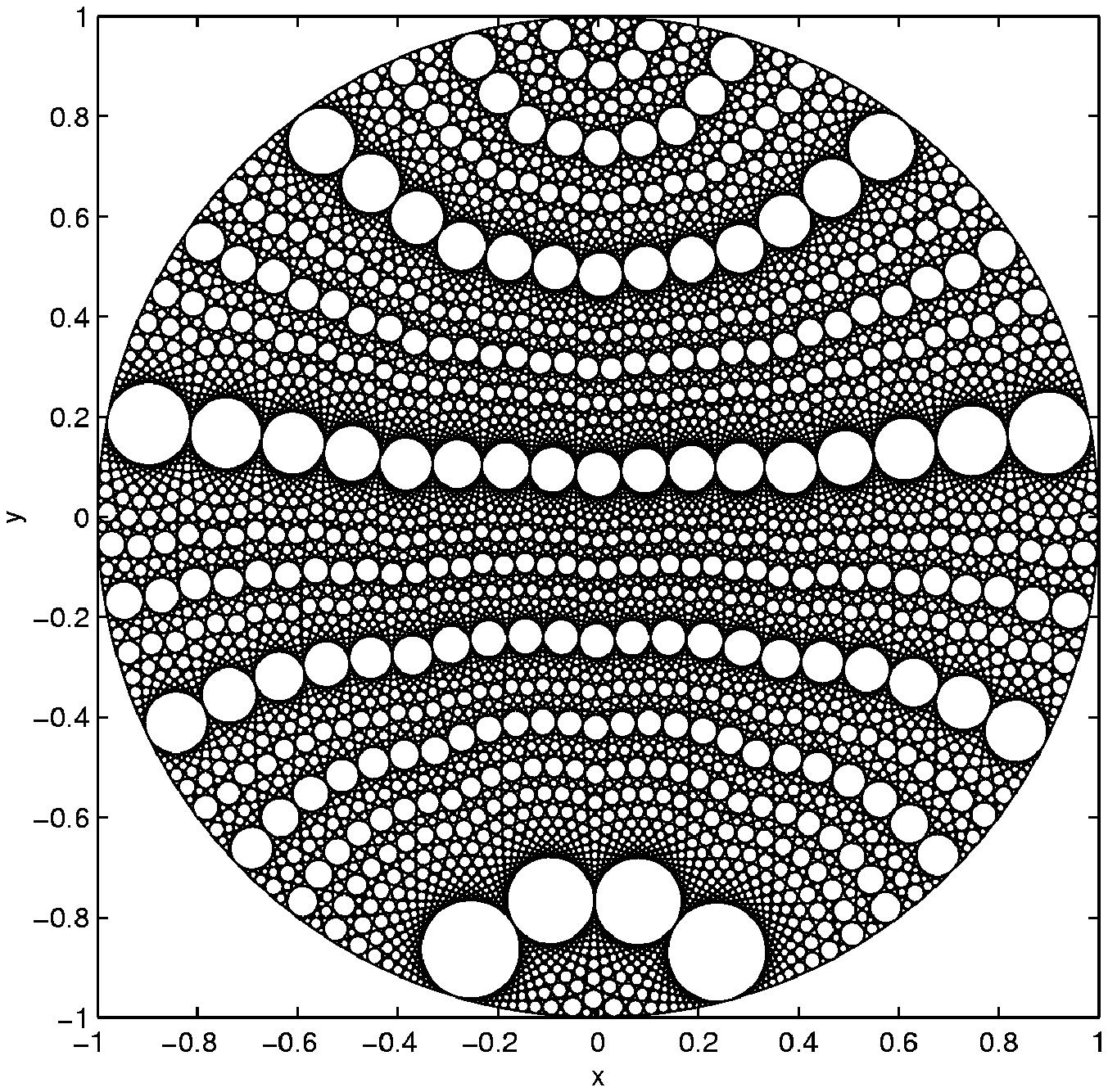}
\caption{The phase space portrait for the spherical map when $\omega=\pi(\sqrt{5}-1)$ and $\mu=0.02$.}
\label{fig9}
\end{figure}

We are left with the question as to whether the resonances of the stable 
periodic orbits densely pack the phase space. Recall that if $\omega/2\pi$ was 
rational in the planar map then the phase plane was found to be tiled with 
polygons. There exist trivial examples of the tiling when 
$\omega/2\pi=0,1/2,1/4,3/4,1/3,2/3,1/6,5/6$. In these cases there are no 
periodic orbits of higher order. When $\omega/2\pi=1/4,3/4$ the phase plane is 
tiled with a grid of squares. A similar situation occurs when 
$\omega/2\pi=1/3,2/3,1/6,5/6$. Then the phase plane is tiled with triangles and 
hexagons. In all of these cases the unstable set (closure of the set of all 
images and preimages of the vertical line $x=0$) has zero Lebesgue measure and 
is one dimensional. For other rationals something different occurs. When 
$\omega/2\pi=1/8,3/8,5/8,7/8$ the phase plane is tiled with octagons of 
decreasing size in a selfsimilar fashion (see Fig. \ref{fig4}). Using selfsimilarity one 
can show that the unstable set has zero measure with the Hausdorff dimension 
\cite{hausdorff} of $\log{3}/\log(1+\sqrt{2})=1.246..$. When 
$\omega/2\pi=1/5,2/5,3/5,4/5,1/10,3/10,7/10,9/10$ the phase plane is tiled with 
pentagons and decagons (see Fig. \ref{fig5}) and the unstable set has the Hausdorff 
dimension of $\log{6}/\log(2+\sqrt{5})=1.241..$. For other rationals the phase 
plane is tiled with many different types of polygons in patterns of great 
complexity (see Fig. \ref{fig11}). We were unable to find the dimension of the unstable 
set for other rationals, but it would be safe to conjecture that they also have 
zero measure. Hence one could assume that whenever $\omega/2\pi$ is rational 
the unstable set has empty interior and the phase plane is densely packed with 
polygonal resonances. However when $\omega/2\pi$ is irrational we are 
uncertain as to whether the phase plane will be densely packed with circles. 
But this seems likely since any irrational can be approximated by a sequence 
of rationals. Hence a phase plane portrait for $\omega/2\pi$ irrational can be 
approximated by a sequence of dense polygonal packings. As we get closer to 
the irrational new smaller polygons are created and the number of sides of 
each existing polygon increases until they approximate circles. This argument 
is quite naive however and a mathematical proof is needed. 

\begin{figure}[p]
\includegraphics[scale=1]{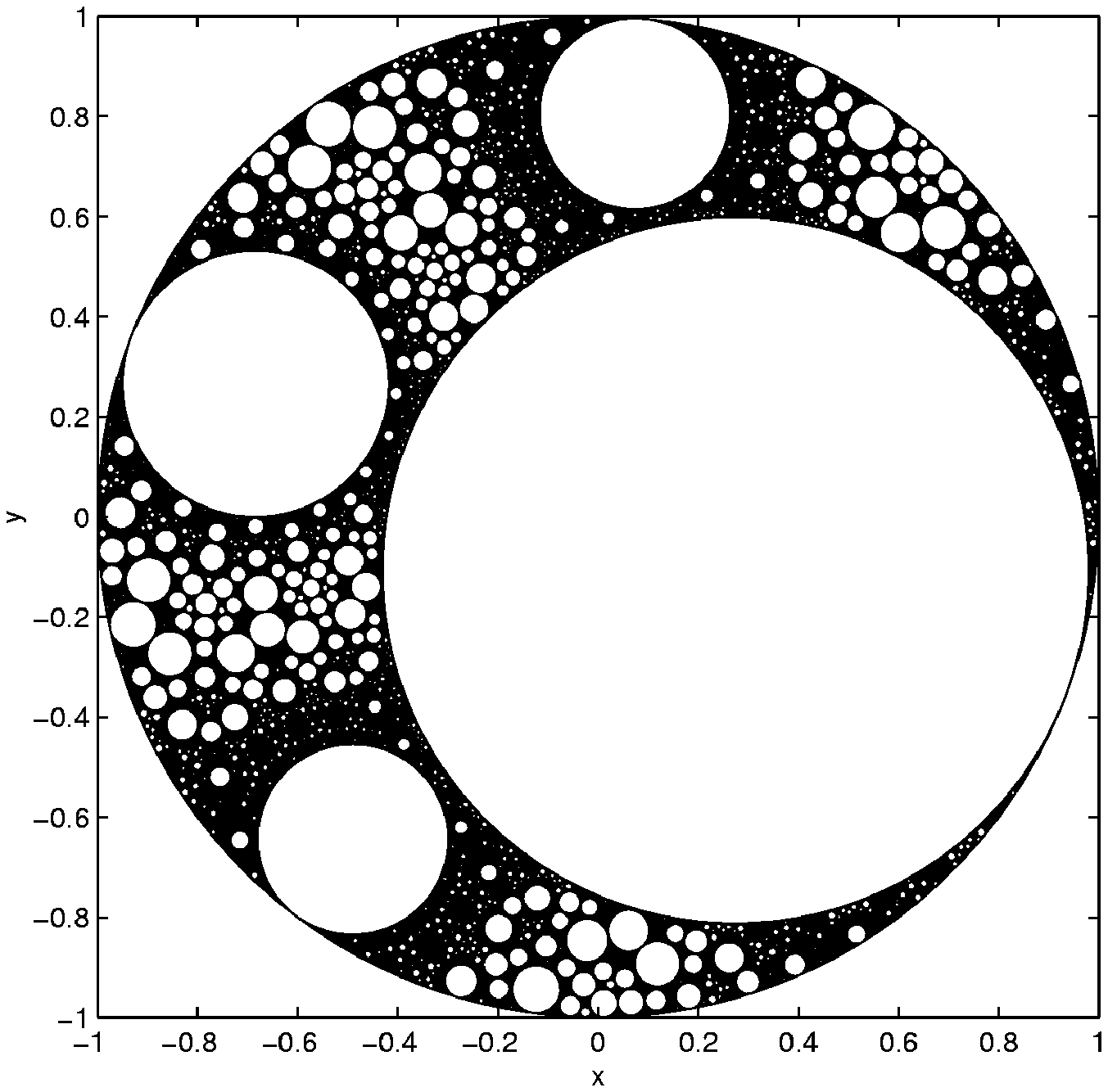}
\caption{The phase space portrait for the spherical map when $\omega=\pi(\sqrt{5}-1)$ and $\mu=\pi(\sqrt{5}-1)$.}
\label{fig10}
\end{figure}

\begin{figure}[p]
\includegraphics[scale=1]{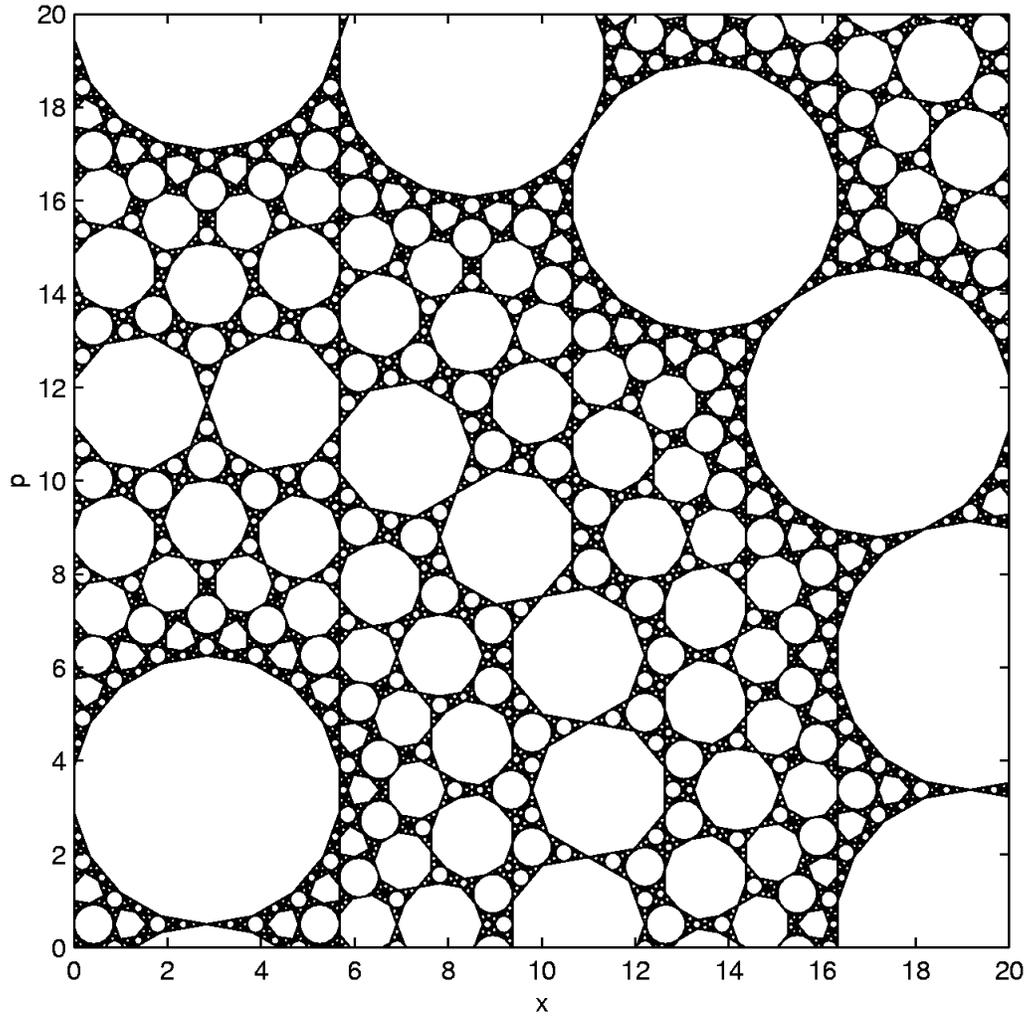}
\caption{Detail of the phase plane when $\omega/2\pi=2/9$ and $\mu=1$ of the planar map.}
\label{fig11}
\end{figure}

\begin{figure}[p]
\includegraphics[scale=1]{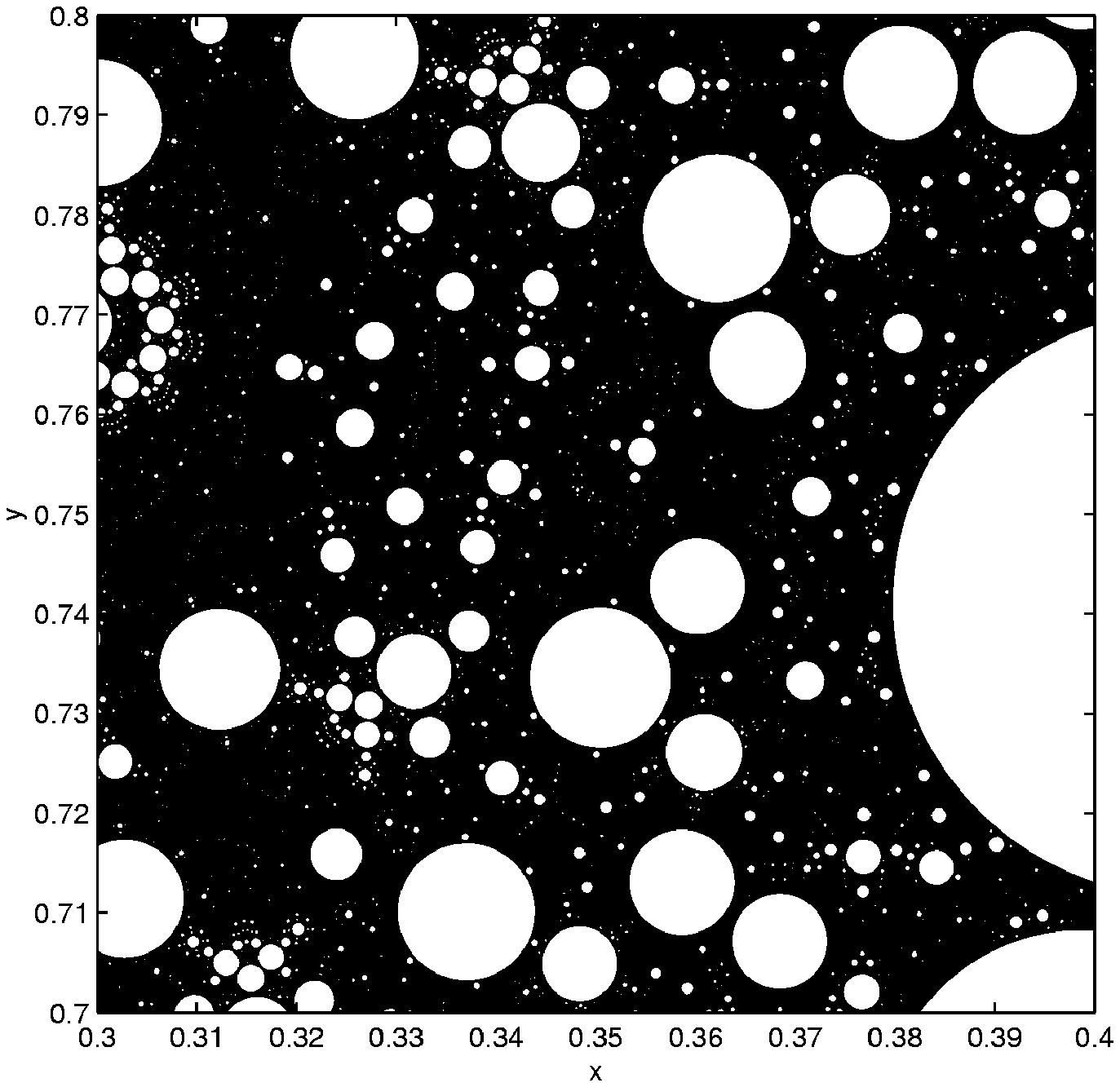}
\caption{Detail of the phase space in Fig. \ref{fig10}.}
\label{fig12}
\end{figure}

In the hyperbolic and spherical cases there are no polygonal packings and the above
argument does not apply. Whether these mappings produce circle packings is also an open 
question. Consider the case when $\omega=\mu=\pi(\sqrt{5}-1)$ in the spherical 
map (Fig. \ref{fig10}). Only the first order periodic orbits of periods 1, 2 and 4 exist. 
The largest circle is a resonance of a period 2 orbit, the second largest is of 
period 1, while the two smaller circles are of period 4. All smaller resonances 
are of higher order. Part of the phase portrait is shown in detail in Fig. \ref{fig12}. 
At this level of magnification one can see that the layout of the high order 
resonances is highly irregular and there seems to be no selfsimilarity. This is 
a great departure from the simple polygonal packings in Fig. \ref{fig4} and Fig. \ref{fig5} where 
the selfsimilarity is immediately apparent. One can also see that if the 
unstable set has zero measure then its fractal dimension would be extremely 
close to two. Indeed, in Fig. \ref{fig12} the unstable set looks more likely to be a set 
of positive measure.

It is worthwhile to point out that phase space structures similar to that in 
the planar mapping can be produced in the sawtooth standard map
\begin{eqnarray*}
x^{n+1}&=&x^n+k\saw(y^n) \\
y^{n+1}&=&y^n+x^{n+1}
\end{eqnarray*}
if $k=-4\sin^2(\theta/2)$, where $(x,y)\in[0,1)^2$ and 
$\saw(y)=y-\lfloor y\rfloor-1/2$. Numerical investigations have led Ashwin 
\cite{ashwin} to conjecture that the closure of the set of images of the 
discontinuity ($y=0$) has positive Lebesgue measure whenever $\theta/\pi$ is 
irrational. If this conjecture proves true we also expect it to hold for our 
planar map (when  $\omega/2\pi$ is irrational) as the two mappings seem closely 
related. We should also point out that the polygonal tilings of the planar map 
also arise in `polygonal dual billiards' 
\cite{vivaldi,tabachnikov1,tabachnikov2}.

Regardless of whether we have discovered new examples of circle packings, 
these deceptively simple mappings demonstrate that phase space structure of 
stunning complexity can arise in Hamiltonian dynamics. As we have shown, a 
large class of periodic orbits can be found. Perhaps some form of periodic 
orbit quantization might be realizable. If nothing else, their aesthetic beauty 
is an admirable quality.



\begin{thebibliography}{9}

\bibitem{bullet}Bullet S and Mantica G 1992 Group theory of hyperbolic circle packings {\it Nonlinearity} {\bf 5} 1085-109.
\bibitem{bullet2}Bullet S and Penrose C 1999 Perturbing circle-packing Kleinian groups as correspondences {\it Nonlinearity} {\bf 12} 635-72.
\bibitem{herrmann}Herrmann H J, Mantica G and Bessis D 1990 Space-filling bearings {\it Phys. Rev. Lett.} {\bf 65} 3223-6.
\bibitem{parker}Parker J R 1995 Kleinian circle packings {\it Topology} {\bf 34} 489-96.
\bibitem{keen}Keen L, Maskit B and Series C 1993 Geometric finiteness and uniqueness for Kleinian groups with circle packing limit sets {\it J. Reine Angew. Math.} {\bf 436} 209-19.
\bibitem{maskit}Maskit B 1988 {\it Kleinian Groups} (Berlin:Springer).
\bibitem{pina}Pi\~{n}a E and Jim\'{e}nez Lara L 1987 On the symmetry lines of the standard mapping {\it Physica} {\bf 26D} 369-78.
\bibitem{lamb}Lamb J S W and Roberts J A G 1998 Time-reversal symmetry in dynamical systems: A survey {\it Physica} {\bf 112D} 1-39.
\bibitem{balazs}Balazs N L and Voros A 1986 Chaos on the pseudosphere {\it Physics Reports} {\bf 143} 109-240.
\bibitem{ratcliffe}Ratcliffe J G 1994 {\it Foundations of Hyperbolic Manifolds} (New York:Springer).
\bibitem{hausdorff}Hausdorff 1918 Dimension und Au{\ss}eres Ma{\ss} {\it Math. Annalen.} {\bf 79} 157.
\bibitem{ashwin}Ashwin P 1997 Elliptic behaviour in the sawtooth standard map {\it Phys. Lett. A} {\bf 232} 409-16.
\bibitem{vivaldi}Vivaldi F and Shaidenko A V 1987 Global stability of a class of discontinuous dual billiards {\it Commun. Math. Phys.} {\bf 110} 625-40.
\bibitem{tabachnikov1}Tabachnikov S 1993 Dual billiards {\it Russ. Math. Surv.} {\bf 48} 75-102.
\bibitem{tabachnikov2}Tabachnikov S 1995 On the dual billiard problem {\it Adv. Math.} {\bf 115} 221-49.


\end{thebibliography}
\end{document}